\documentclass[sigconf]{acmart}

\AtBeginDocument{%
  }

\usepackage{booktabs}
\usepackage{multirow}
\usepackage{enumitem}
\usepackage{xcolor}
\usepackage{xspace}
\usepackage{csquotes}
\usepackage{tikz}
\usetikzlibrary{positioning,arrows.meta,fit,backgrounds,calc,shapes.geometric}
\graphicspath{{fig/}}

\newcommand{\method}{CAMIE}

\usepackage[normalem]{ulem}

\setcopyright{none}
\copyrightyear{2027}
\acmYear{2027}
\acmDOI{}
\acmConference[KDD '27, ADS Track, Under Review]{The 33rd ACM SIGKDD Conference on Knowledge Discovery and Data Mining}{August 1--5, 2027}{San Jose, CA, USA}

\acmISBN{}
\begin{document}

\title[CAMIE]{CAMIE: Co-Engagement-Aware Multimodal Item Embeddings for Snap Dynamic Product Ads Retrieval}

\author{Xiaodong Liu}
\affiliation{%
  \institution{Snap Inc.}
  \city{Bellevue}
  \state{WA}
  \country{USA}
}
\email{xliu9@snapchat.com}

\author{Siman Wang}
\affiliation{%
  \institution{Snap Inc.}
  \city{Bellevue}
  \state{WA}
  \country{USA}
}
\email{swang7@snapchat.com}

\author{Congfei Zhang}
\affiliation{%
  \institution{Snap Inc.}
  \city{Bellevue}
  \state{WA}
  \country{USA}
}
\email{czhang3@snapchat.com}

\author{Hsiang-wei Chao}
\affiliation{%
  \institution{Snap Inc.}
  \city{Seattle}
  \state{WA}
  \country{USA}
}
\email{hchao@snapchat.com}

\author{Xiao Bai}
\affiliation{%
  \institution{Snap Inc.}
  \city{Palo Alto}
  \state{CA}
  \country{USA}
}
\email{xbai@snapchat.com}

\author{Wen Zhang}
\affiliation{%
  \institution{Snap Inc.}
  \city{Bellevue}
  \state{WA}
  \country{USA}
}
\email{wzhang7@snapchat.com}

\author{Jingxiao Ma}
\affiliation{%
  \institution{Snap Inc.}
  \city{Bellevue}
  \state{WA}
  \country{USA}
}
\email{jma3@snapchat.com}

\author{Zhe Liu}
\affiliation{%
  \institution{Snap Inc.}
  \city{Palo Alto}
  \state{CA}
  \country{USA}
}
\email{zliu11@snapchat.com}

\author{Yunzhi Zhou}
\affiliation{%
  \institution{Snap Inc.}
  \city{Palo Alto}
  \state{CA}
  \country{USA}
}
\email{yzhou10@snapchat.com}

\author{Yajun Wang}
\affiliation{%
  \institution{Snap Inc.}
  \city{Palo Alto}
  \state{CA}
  \country{USA}
}
\email{ywang30@snapchat.com}

\author{Jinchao Li}
\affiliation{%
  \institution{Snap Inc.}
  \city{Bellevue}
  \state{WA}
  \country{USA}
}
\email{jli18@snapchat.com}

\author{Yu Zhang}
\affiliation{%
  \institution{Snap Inc.}
  \city{Palo Alto}
  \state{CA}
  \country{USA}
}
\email{yzhang3@snapchat.com}

\renewcommand{\shortauthors}{Liu et al.}

\begin{abstract}
Item-to-item (I2I) retrieval is a core primitive in large-scale
recommendation and advertising systems. In production Snap Dynamic
Product Ads (DPA), I2I retrieval faces two challenges:
separate visual, textual, and multimodal encoders fragment the
retrieval stack, and content-only training does not align
embeddings with the co-engagement behavior that drives downstream
conversions. We present \textbf{\method{}}, a co-engagement-aware
multimodal item embedding framework for Snap DPA retrieval.
\method{} builds on LLM/MLLM backbones, using their native
multimodal interfaces to represent
item images and metadata in a shared embedding space. It then
fine-tunes the backbone on co-engaged item pairs mined from user
journeys with a symmetric in-batch InfoNCE objective. Offline,
\method{} outperforms the strongest commercial multimodal
embedding model on Recall@10 and serves text-only retrieval from
the same checkpoint with minimal quality loss. Online, \method{} serves
as a drop-in replacement for two deployed content-based I2I encoders,
delivering $+0.390\%$ CTR\,/\,$+10.832\%$ CVR over the multimodal
control, $+18.958\%$ CTR\,/\,$+13.12\%$ CVR over the text control, and
$+0.211\%$ CTR\,/\,$+1.911\%$ CVR on overall DPA traffic. \method{} is deployed in production.
\end{abstract}


% \ccsdesc[300]{Computing methodologies~Learning latent representations}


\maketitle

\section{Introduction}
\label{sec:intro}

Item-to-item (I2I) retrieval is a core candidate-generation
primitive in recommendation and advertising
systems~\cite{linden2003amazon}.
Given a seed item, an I2I retriever returns related catalog items.
Production systems typically encode the catalog offline, build an
approximate-nearest-neighbor (ANN)~\cite{guo2020scann,
johnson2019faiss} index, and answer each request with a single
index lookup. This design keeps online latency low, scales to
large catalogs, and naturally covers cold-start and long-tail items
that two-tower personalized
retrievers~\cite{huang2013dssm,covington2016youtubednn,yi2019sampling}
struggle to serve.

Two constraints limit content-based production I2I. First,
\textbf{modality fragmentation}: different item modalities capture
different notions of similarity. Images are useful for visual
substitutes, text and category metadata capture product semantics
and catalog structure, and multimodal encoders combine both. In
practice this leads to multiple coexisting retrieval stacks:
text-only encoders, image-only encoders, and multimodal encoders
(commonly CLIP/SigLIP variants~\cite{radford2021clip,zhai2023siglip,tschannen2025siglip2}),
each with its own fine-tune, refresh pipeline, and ANN index. What
is missing is a single multimodal item encoder that can also serve
single-modality use cases when only text or only an image is
available.

Second, \textbf{content-only supervision}: most content-based I2I
models learn from item metadata itself, including image, title,
description, brand, and category, so their embedding space is
optimized for content similarity rather than the engagement space
used by the recommender system. User-journey logs, however, provide
a dense behavioral signal: users repeatedly view, click, add to cart, and
purchase related items within coherent journeys. Ignoring these
co-engagement events leaves the retriever weakly aligned with the
downstream objective. The two problems compound: production systems
maintain separate modality-specific encoders, and those encoders
are usually not trained on the journey signal most predictive of
what users will engage with next.

Behavior-aware LLM/MLLM retrieval
methods~\cite{zhang2024notellm,zhang2024notellmv2} show that
engagement signals can adapt pretrained representations, but they
do not target a drop-in production I2I encoder serving multimodal,
text-only, and image-only retrieval. This setting motivates a general-purpose embedding model with
native multimodal inputs and shared attention over item fields,
fine-tuned to reflect Snap DPA co-engagement patterns. It also raises
an attribution question that prior work leaves open: how much of the
resulting quality comes from the pretrained backbone, and how much
from the co-engagement supervision? We answer it directly by training
purpose-built non-LLM dual encoders under an identical recipe, and
find that the supervision accounts for most of the gain while the
MLLM backbone adds a small but significant multimodal advantage and a
large text-only advantage.

We present \method{}, a co-engagement-supervised LLM/MLLM
embedding framework for industrial I2I retrieval in Snap Dynamic
Product Ads (DPA). \method{} renders item images and metadata as a
single multimodal record, fine-tunes an embedding backbone on
co-engaged item pairs mined from user journeys, and serves the
resulting vectors through the existing ANN retrieval path. The same
trained checkpoint is designed to support multimodal, text-only,
and image-only I2I retrieval. Our contributions are:
\begin{enumerate}[leftmargin=*,nosep]
\item We propose \method{}, a
  co-engagement-supervised LLM/MLLM embedding framework for
  industrial I2I retrieval. A single multimodal backbone is
  fine-tuned on co-engaged item pairs with symmetric in-batch
  InfoNCE and serves multimodal, text-only, and image-only
  retrieval from one checkpoint.
\item On a held-out Snap
  DPA test set, \method{} outperforms the strongest commercial
  multimodal embedding model on Recall@10, and the same checkpoint
  serves text-only retrieval at near-parity with a dedicated text-only
  fine-tune. A matched-recipe comparison against purpose-built dual
  encoders credits most of the gain to the co-engagement supervision
  rather than to the backbone, with image-only retrieval the exception. To our knowledge this is the
  first controlled attribution of this kind for a deployed multimodal
  I2I encoder.
\item \method{} is deployed in
  Snap DPA as a drop-in replacement for two content-based I2I
  encoders, with $+0.390\%$ CTR /
  $+10.832\%$ CVR over the multimodal
  control and $+18.958\%$ CTR /
  $+13.12\%$ CVR over the text control
  in online A/B tests, and $+0.211\%$ CTR / $+1.911\%$ CVR on
  overall DPA traffic, demonstrating effectiveness in a
  production recommendation system.
\end{enumerate}

\section{Related Work}
\label{sec:related}

\noindent\textbf{Content-based I2I.}
Classical I2I estimates similarity from co-purchase or co-occurrence
counts~\cite{linden2003amazon}. These methods are
simple and scalable but rely on historical interactions, limiting
cold-start coverage. Content-based I2I improves coverage by
representing items from metadata, commonly with
CLIP~\cite{radford2021clip} or SigLIP/SigLIP2
backbones~\cite{zhai2023siglip,tschannen2025siglip2}.

\noindent\textbf{Industrial multimodal and unified item embeddings.}
The closest industrial systems learn a \emph{single} item
representation shared across surfaces.
ItemSage~\cite{baltescu2022itemsage} aggregates text and image
features with a transformer to produce one set of Pinterest product
embeddings serving user-, image-, and search-based recommendations,
and uses multi-task learning over several engagement types.
CommerceMM~\cite{yu2022commercemm} pre-trains a commerce multimodal
model with omni-retrieval tasks spanning text-, image-, and
multimodal-to-multimodal mappings.
OmniSearchSage~\cite{pancha2024omnisearchsage} jointly learns query,
pin, and product embeddings for Pinterest search, with compatibility
constraints to pre-existing embeddings.
MERLIN~\cite{tiady2024merlin} builds multimodal, multilingual
recommendations from a product graph of mined item associations using
a GNN with dual embeddings and cold-start metadata.
\method{} shares the goal of one unified item representation, but
differs in three ways: the encoder is an MLLM consumed through its
native multimodal interface rather than a fusion module over frozen
unimodal features; supervision comes from co-engagement pairs mined
from DPA journeys rather than content or graph associations; and a
single checkpoint is evaluated \emph{and deployed} across multimodal,
text-only, and image-only serving.

\noindent\textbf{LLM/MLLM-based retrieval.}
NoteLLM~\cite{zhang2024notellm} compresses a note into a single
\texttt{[EMB]} token with a co-engagement contrastive loss, and
NoteLLM-2~\cite{zhang2024notellmv2} extends this idea to a
multimodal backbone with a visual contrastive routing loss.
Both target note recommendation and introduce architectural
machinery (a compression token, a visual routing loss) to fuse
modalities; \method{} keeps the backbone's native multimodal
interface unchanged and instead targets a drop-in production I2I
encoder that serves multimodal, text-only, and image-only traffic
from one checkpoint.
LLM-I2I~\cite{feng2025llmi2i} and LLaRA~\cite{liao2024llara} use
LLMs for augmentation or reasoning rather than as the serving item
encoder. \method{} instead studies LLM/MLLM backbones as the
serving item encoder for Snap DPA: pretrained world knowledge gives
the model a broad product representation across modalities, and
co-engagement fine-tuning aligns that representation with
production DPA behavior. The resulting embeddings are served through
a standard ANN index, and one checkpoint is evaluated for
multimodal, text-only, and image-only retrieval.

\section{Method}
\label{sec:method}

\subsection{Overview}
\label{sec:overview}

\begin{figure}[t]
  \centering
  \resizebox{0.95\columnwidth}{!}{%
  \begin{tikzpicture}[
    >={Stealth[length=2mm]},
    font=\small,
    box/.style={draw, rounded corners=2pt, align=center,
                inner sep=4pt, minimum height=7mm},
    inp/.style={box, fill=gray!8,    draw=gray!60,   minimum width=32mm},
    enc/.style={box, fill=purple!10, draw=purple!60, minimum width=72mm},
    emb/.style={draw=blue!55, thick, circle, inner sep=1.5pt, fill=blue!8,
                font=\small},
    loss/.style={box, fill=green!10, draw=green!60,  minimum width=72mm},
    pip/.style={->, semithick, draw=black!55}
  ]

  \node[inp] (xa) at (-20mm, 0)
       {Item $i$\\\scriptsize optional image $+$ title $+$ brand\\\scriptsize $+$ description $+$ category};
  \node[inp] (xb) at ( 20mm, 0)
       {Item $j$\\\scriptsize optional image $+$ title $+$ brand\\\scriptsize $+$ description $+$ category};

  \node[enc, below=14mm of $(xa)!0.5!(xb)$] (bb)
       {LLM/MLLM embedding backbone\\\scriptsize optional adapter fine-tune};
  \node[enc, below=4mm of bb] (head) {Projection head};
  \node[enc, below=4mm of head] (norm) {$L_2$ normalize};

  \node[emb, below=4mm of norm.south, xshift=-20mm] (za) {$z_i$};
  \node[emb, below=4mm of norm.south, xshift= 20mm] (zb) {$z_j$};

  \node[loss, below=12mm of norm.south, align=center] (loss)
       {Symmetric in-batch InfoNCE};

  \draw[pip] (xa.south) -- (xa.south |- bb.north);
  \draw[pip] (xb.south) -- (xb.south |- bb.north);
  \draw[pip] (bb)   -- (head);
  \draw[pip] (head) -- (norm);
  \draw[pip] (norm.south -| za) -- (za);
  \draw[pip] (norm.south -| zb) -- (zb);
  \draw[pip] (za.south) -- (za.south |- loss.north);
  \draw[pip] (zb.south) -- (zb.south |- loss.north);

  \end{tikzpicture}%
  }
  \caption{The \method{} framework. Two items from a
  co-engagement pair are rendered with the fields available for the
  target serving mode, encoded by a shared LLM/MLLM embedding
  backbone, passed through a projection head, and
  $L_2$-normalized. Training uses symmetric in-batch InfoNCE with
  in-batch and cross-device negatives.}
  \Description{Schematic of the CAMIE framework. Two item inputs,
  each containing an optional image plus title, brand, description,
  and category text, are fed into a shared LLM/MLLM embedding
  backbone with an optional adapter fine-tune. A projection head and $L_2$
  normalization produce two embeddings, which feed into a symmetric
  in-batch InfoNCE loss.}
  \label{fig:framework}
\end{figure}
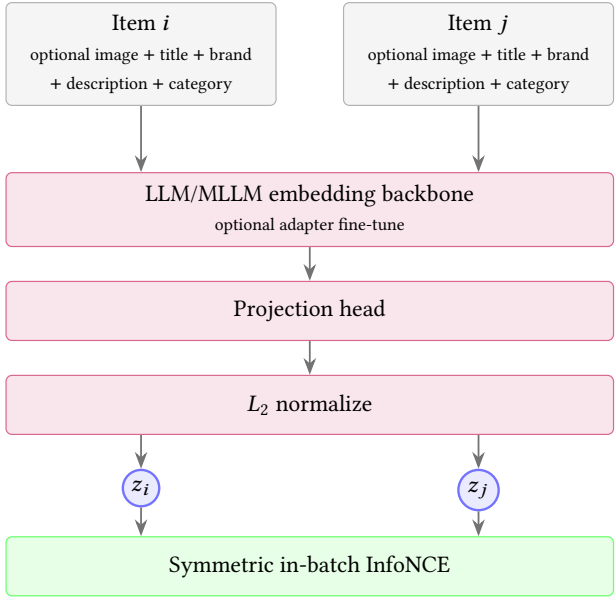

\method{} aligns a pretrained LLM/MLLM embedding backbone with
item co-engagement behavior. The pipeline has four steps:
(1) mine unordered positive item pairs from user journeys;
(2) render each item as a modality-conditioned record; (3) encode
both items with a shared backbone and trainable projection head;
and (4) optimize the embeddings with a symmetric in-batch
contrastive objective, then serve them through a standard ANN
index. Figure~\ref{fig:framework} summarizes steps (2)--(4).
The framework is backbone-agnostic; we focus on LLM/MLLM backbones
because their native multimodal inputs and pretrained knowledge of
products, brands, attributes, and categories motivate the
modality-unified design.
Section~\ref{sec:datasets} describes the concrete Snap DPA
instantiation used in our experiments.

\subsection{Co-engagement pair construction}
\label{sec:pair-construction}

We construct supervision as a graph of co-engaged item pairs. Let
a user journey be a time-ordered sequence of engagement events
$(u, i, t, a)$, where user $u$ interacts with item $i$ at time $t$
through action type $a$ (for example, view, click, add-to-cart, or
purchase). A pair miner maps journeys to unordered positive edges
\begin{equation}
  \mathcal{D}
  = \{(i, j, \omega_{ij}) : \exists u,\ i \ne j,\
       i,j \in \mathcal{J}_u,\
       \mathrm{scope}(i)=\mathrm{scope}(j)\},
  \label{eq:pair-set}
\end{equation}
where $\mathcal{J}_u$ is the set of items observed in a local
session or attribution window for user $u$, $\mathrm{scope}(\cdot)$
denotes the serving scope in which retrieval is allowed, and
$\omega_{ij} \in (0,1]$ is an optional reliability weight.

\method{} does not require positives to come only from high-intent
events. The pair miner can use the full engagement funnel and apply
quality filters for attribution confidence, required item fields,
serving-scope consistency, temporal proximity, and taxonomy
coherence. Temporal weights, when enabled, down-weight distant
co-engagements with an exponential half-life; the unweighted
objective is the special case $\omega_{ij}=1$. Concrete filters and
thresholds are data-dependent and are reported with the
experimental setup (Section~\ref{sec:datasets}).

\subsection{Modality-conditioned item encoder}
\label{sec:encoder}

\noindent\textbf{Item input format.}
Each item $i$ can expose a primary image $\mathcal{I}_i$, title
$t_i$, brand $b_i$, free-text description $d_i$, and product
category path $c_i$. A renderer $R_m$ converts those fields into
the input sequence for serving mode
$m \in \{\text{multimodal}, \text{text-only}, \text{image-only}\}$:
\begin{equation}
\begin{aligned}
  x_i^{(m)} = R_m\bigl(&\mathcal{I}_i,\;
                  \texttt{"Title: "} t_i,\;
                  \texttt{"Brand: "} b_i,\;\\
                 &\texttt{"Description: "} d_i,\;
                  \texttt{"Category: "} c_i\bigr).
\end{aligned}
  \label{eq:item-input}
\end{equation}
For multimodal serving, the renderer includes both image and text
fields. For text-only serving, it omits the image; for image-only
serving, it omits the text fields. The image, when present, is
processed by the backbone's native visual input pipeline, and text
fields are processed by the backbone tokenizer. No modality-specific
projection head is required.

\noindent\textbf{Shared encoder and projection.}
Let $f_\theta$ denote the shared embedding backbone and $g_\phi$
the projection head used for serving. The backbone may be a
decoder-only multimodal embedding model, a text-only LLM embedding
model, or another embedding-capable architecture. Decoder-only
models use last-token pooling; models with native pooling can use
their own pooling head. The final serving vector is
\begin{equation}
  z_i^{(m)}
  = \frac{g_\phi(f_\theta(x_i^{(m)}))}
         {\|g_\phi(f_\theta(x_i^{(m)}))\|_2}
  \in \mathbb{R}^{d}.
  \label{eq:item-embedding}
\end{equation}
The same $f_\theta$ and $g_\phi$ are used for all serving modes,
so changing modality changes only which fields are rendered in
$x_i^{(m)}$.

\subsection{Symmetric in-batch InfoNCE}
\label{sec:loss}

In distributed training, embeddings are all-gathered across devices,
so the denominator uses both local and cross-device negatives. Let
$\mathcal{B}=\{(i_n,j_n,\omega_n)\}_{n=1}^{N}$ denote the gathered
batch of $N$ co-engagement pairs. The weighted symmetric
InfoNCE~\cite{oord2018cpc} objective is
\begin{equation}
\begin{aligned}
  \mathcal{L}
  = -\frac{1}{2\sum_{n=1}^{N}\omega_n}
  \sum_{n=1}^{N}\omega_n \Bigg[
  &\log\frac{\exp(z_{i_n}^{\top}z_{j_n}/\tau)}
             {\sum_{k=1}^{N}\exp(z_{i_n}^{\top}z_{j_k}/\tau)}
  \\
  +&\log\frac{\exp(z_{j_n}^{\top}z_{i_n}/\tau)}
             {\sum_{k=1}^{N}\exp(z_{j_n}^{\top}z_{i_k}/\tau)}
  \Bigg],
\end{aligned}
  \label{eq:infonce}
\end{equation}
where $\tau$ is a temperature parameter. Symmetry treats
$(i,j)$ and $(j,i)$ as the same positive relation, which matches
unordered item co-engagement. In the unweighted configuration,
all $\omega_n$ are set to $1$.

\subsection{ANN serving}
\label{sec:serving}

\noindent\textbf{Index build.}
At each catalog refresh, every active item is rendered under the
target index modality, encoded once by $f_\theta$ and $g_\phi$,
normalized, and written to an ANN index. The embedding dimension,
index type, and quantization settings are deployment choices; the
framework requires only that the same item representation used in
training can be materialized offline for retrieval.

\noindent\textbf{Online retrieval.}
At request time, a seed item is served by identifier lookup when
its embedding is already cached, or by encoding a fresh query record
under the appropriate modality. The request then issues a single
ANN top-$K$ query. Because the embedding backbone runs at
index-build time for catalog items, the online path keeps the same
latency shape as conventional content-based I2I retrieval.

\noindent\textbf{Modality-conditioned serving.}
The same trained checkpoint can support multimodal, text-only, and
image-only surfaces by changing the renderer $R_m$ while holding
the encoder and index interface fixed. Section~\ref{sec:modality-flexibility}
measures the quality cost of those modality-restricted serving
modes in the Snap DPA instantiation.

\section{Experiments}
\label{sec:experiments}

We evaluate \method{} on a fixed Snap DPA train/test split. The
section is organized around four questions:
\textbf{(Q1)} How much does co-engagement fine-tuning improve over
pretrained backbones and commercial APIs?
\textbf{(Q2)} Can one fine-tuned checkpoint serve multimodal,
text-only, and image-only retrieval?
\textbf{(Q3)} Does training-data scale matter more than filtering
for high-intent events?
\textbf{(Q4)} How much of the gain comes from the LLM/MLLM
backbone as opposed to the co-engagement supervision itself?

\subsection{Experimental setup}
\label{sec:datasets}

\noindent\textbf{Data.}
We instantiate the generic pair miner from Section~\ref{sec:pair-construction}
on an internal $30$-day Snap DPA user-journey table. We deduplicate
each user's journey by $\texttt{item\_id}$, compute pairwise
temporal-decay weights with a $7$-day half-life, and sample up to
$2$ pairs per user. We keep page views, clicks, add-to-carts, and
purchases, and remove noisy pairs by requiring \texttt{EXACT} match
attribution, non-empty title and image fields, the same advertiser,
decay weight at least $0.5$, and coherent top-level category. The
offline split used for all tables in this section contains $10$M
training pairs and $100$K test pairs; the deployed checkpoint in
Section~\ref{sec:online} is trained on a larger $100$M-pair
production set built with the same pipeline.

\noindent\textbf{Metrics and protocol.}
Offline metrics are Recall@K for $K \in \{1,10,50\}$ and MRR. For
each test pair, the query item is scored against the candidate pool
of all unique items appearing in the test pairs. Online metrics are
relative CTR and CVR changes, reported under Snap's disclosure
policy.

\noindent\textbf{Implementation details.}
Our primary LLM/MLLM instantiation uses
Qwen3-VL-Embedding 2B~\cite{qwen3vlembedding}. We freeze the
backbone, add LoRA adapters~\cite{hu2022lora}
($r{=}32$, $\alpha{=}64$) on $\{q,k,v,o\}$ attention projections,
and train the adapters with the projection head. We use
AdamW~\cite{loshchilov2019adamw} with peak learning rate
$2{\times}10^{-5}$, $200$ warm-up steps, cosine decay, BF16, and a
learnable temperature initialized to $0.07$. Effective batch size
is $32$--$64$ on $8{\times}$ NVIDIA A100 40\,GB GPUs. The primary
checkpoint trains for $1$ epoch over the training set.

Offline tables use $512$-dimensional embeddings. Online serving
uses the first $128$ dimensions so the index footprint matches the
existing \texttt{I2I\_MM} index. All \method{} fine-tunes in
Table~\ref{tab:leaderboard} use the same training pairs.
\textbf{OSS} rows use native pretrained weights; \textbf{API} rows
use commercial API inference at default configuration.

\subsection{Q1: Model comparison}
\label{sec:leaderboard}

We first compare the Qwen3-VL instantiation of \method{} with
pretrained open-source checkpoints (OSS) and commercial embedding
APIs (API).

\begin{table*}[t]
  \caption{Model comparison on the held-out test set. Scores are
  normalized to the pretrained Qwen3-VL-Embedding 2B multimodal
  row; higher is better. \textbf{\method{}} rows are fine-tuned on
  the training set; \textbf{OSS} rows are pretrained open-source
  checkpoints; \textbf{API} rows are commercial embedding services.
  Best per column is bold.}
  \label{tab:leaderboard}
  \small
  \setlength{\tabcolsep}{4pt}
  \begin{tabular}{lllcccc}
    \toprule
    Model                  & Type                & Input              & R@1                & R@10               & R@50               & MRR                \\
    \midrule
    Qwen3-VL-Embedding 2B  & \textbf{\method{}}  & Multimodal         & $126.4\%$          & $\mathbf{137.7\%}$ & $\mathbf{141.8\%}$ & $\mathbf{130.5\%}$ \\
    Qwen3-VL-Embedding 2B  & OSS                 & Multimodal         & $100.0\%$          & $100.0\%$          & $100.0\%$          & $100.0\%$          \\
    Qwen3-VL-Embedding 2B  & \textbf{\method{}}  & Text-only          & $117.8\%$          & $122.5\%$          & $134.1\%$          & $116.4\%$          \\
    Qwen3-VL-Embedding 2B  & OSS                 & Text-only          & $113.2\%$          & $111.1\%$          & $109.7\%$          & $111.9\%$          \\
    \midrule
    Gemini Embedding 2     & API                 & Multimodal         & $\mathbf{129.1\%}$ & $129.9\%$          & $129.2\%$          & $129.4\%$          \\
    Gemini Embedding 2     & API                 & Text-only          & $128.6\%$          & $126.6\%$          & $125.9\%$          & $127.1\%$          \\
    \midrule
    Gemma 4 E2B            & \textbf{\method{}}  & Multimodal         & $94.5\%$           & $108.2\%$          & $124.4\%$          & $100.0\%$          \\
    Gemma 4 E2B            & \textbf{\method{}}  & Text-only          & $93.0\%$           & $103.3\%$          & $120.9\%$          & $97.1\%$           \\
    Gemma 4 E2B            & OSS                 & Text-only          & $93.8\%$           & $95.2\%$           & $97.3\%$           & $93.7\%$           \\
    Gemma 4 E2B-it         & \textbf{\method{}}  & Text-only          & $91.6\%$           & $94.2\%$           & $108.2\%$          & $93.1\%$           \\
    Gemma 4 E2B-it         & OSS                 & Text-only          & $93.7\%$           & $89.6\%$           & $91.3\%$           & $91.5\%$           \\
    \bottomrule
  \end{tabular}
\end{table*}

\noindent\textbf{Co-engagement fine-tuning aligns the pretrained backbone.}
With the same Qwen3-VL-Embedding 2B backbone and the same
multimodal inputs, \method{} reaches $137.7\%$ R@10 versus
$100.0\%$ for the pretrained checkpoint. Fine-tuning also changes
how the model uses images: the pretrained checkpoint is better with
text-only input than with multimodal input ($111.1\%$ vs.\
$100.0\%$ R@10), while the fine-tuned checkpoint benefits from
multimodal input ($137.7\%$ vs.\ $122.5\%$ R@10).

\noindent\textbf{\method{} is stronger than the commercial baselines at
$K \geq 10$.} Gemini Embedding~2 is the strongest
zero-shot baseline at $129.9\%$ R@10. \method{} improves on it by
$\sim$$6\%$ relative while using smaller embeddings ($d{=}512$ vs.\
$d{=}3072$). Gemini stays ahead at $K{=}1$ ($129.1\%$ vs.\
$126.4\%$), so \method{}'s advantage is concentrated at the larger
cut-offs that matter for candidate generation.

\noindent\textbf{Backbone choice still matters.}
Fine-tuned Gemma~4 E2B reaches $108.2\%$ R@10 with multimodal input
and improves over its pretrained checkpoint on the modality they
share ($103.3\%$ vs.\ $95.2\%$ text-only), but remains $29.5$pp below
the Qwen3-VL instantiation. This suggests that the co-engagement framework is
portable, while the starting embedding backbone remains important.

\subsection{Q2: Modality flexibility}
\label{sec:modality-flexibility}

The primary fine-tune is trained with multimodal inputs. The
production fleet, however, also includes text-only and image-only
retrieval surfaces that today consume dedicated unimodal encoders.
We therefore evaluate the same multimodal-trained checkpoint under
three input regimes and compare it with dedicated unimodal
fine-tunes trained on the matching modality
(Table~\ref{tab:modality-flex}).

\begin{table}[t]
  \caption{Modality-restricted serving on the held-out test set.
  All rows use Qwen3-VL-Embedding 2B and the same training pairs.
  Scores are normalized to the multimodal-trained checkpoint served
  with multimodal input; higher is better.}
  \label{tab:modality-flex}
  \small
  \setlength{\tabcolsep}{4pt}
  \begin{tabular}{llcc}
    \toprule
    Training input          & Eval input  & R@10              & MRR               \\
    \midrule
    \multicolumn{4}{l}{\emph{Single multimodal-trained checkpoint:}} \\
    Multimodal              & Multimodal  & $\mathbf{100.0\%}$ & $\mathbf{100.0\%}$ \\
    Multimodal              & Text-only   & $87.9\%$ & $88.7\%$ \\
    Multimodal              & Image-only  & $67.0\%$ & $65.7\%$ \\
    \midrule
    \multicolumn{4}{l}{\emph{Dedicated unimodal fine-tunes:}} \\
    Text-only               & Text-only   & $89.0\%$ & $89.2\%$ \\
    Image-only              & Image-only  & $68.8\%$ & $67.8\%$ \\
    \bottomrule
  \end{tabular}
\end{table}

\noindent\textbf{Text-only serving is effectively unified.}
The multimodal-trained checkpoint served with text-only input
comes within $1.1$pp of a dedicated text-only fine-tune on R@10
($87.9\%$ vs.\ $89.0\%$) and within $0.5$pp on MRR ($88.7\%$ vs.\
$89.2\%$). This supports
replacing the separate text encoder with the same checkpoint used
for multimodal serving.

\noindent\textbf{Image-only serving is limited by the modality, not by
consolidation.}
Image-only serving is markedly weaker than multimodal serving for the
unified checkpoint ($67.0\%$ R@10), but a dedicated image-only
fine-tune is barely better ($68.8\%$, a $1.8$pp gap). Consolidating
onto one checkpoint therefore costs little on this surface; what costs
quality is discarding text at query time. Text appears to be the
dominant co-engagement signal in this catalog: text-only serving stays
within $13$pp of multimodal serving, whereas both image-only
configurations lose roughly a third of multimodal R@10.
Section~\ref{sec:matched-backbones} revisits this surface with matched
non-LLM baselines and shows that a contrastively-trained vision tower
is substantially stronger there than either configuration above,
making a dedicated image encoder the better choice for image-only
traffic.

\subsection{Q3: Training-data scale vs.\ conversion-type purity}
\label{sec:data-scale}

High-intent events provide cleaner positives, but filtering to them
removes many usable pairs. We compare the full all-event training
set with a high-intent subset containing add-to-cart and purchase pairs (Table~\ref{tab:data-scale}). The high-intent
subset is roughly one third of the full training set; the all-event
$1/3$ row is a random downsample matched to that size.

\begin{table}[t]
  \caption{Training-pair scale vs.\ high-intent filtering on the
  held-out test set. R@10 and MRR are normalized to the all-pairs
  $1/3$ baseline; higher is better.}
  \label{tab:data-scale}
  \small
  \setlength{\tabcolsep}{4pt}
  \begin{tabular}{llcc}
    \toprule
    Training filter            & Scale     & R@10     & MRR      \\
    \midrule
    All pairs                  & $1$   & $\mathbf{108.9\%}$ & $\mathbf{107.5\%}$ \\
    High-intent only           & $1/3$ & $101.6\%$           & $101.8\%$           \\
    All pairs                  & $1/3$ & $100.0\%$           & $100.0\%$           \\
    \bottomrule
  \end{tabular}
\end{table}

\noindent\textbf{Scale matters more than event purity.}
At matched scale, high-intent filtering is slightly better than
all-pairs ($+1.6\%$ relative R@10). At matched filter, using the
full all-event training set improves R@10 by $+8.9\%$ relative.
The full all-event training set therefore beats high-intent-only by
$+7.2\%$ relative R@10: the scale gained by keeping more pairs is
larger than the purity gained by filtering.

\subsection{Q4: Backbone vs.\ co-engagement supervision}
\label{sec:matched-backbones}

Sections~\ref{sec:leaderboard}--\ref{sec:modality-flexibility}
compare \method{} against pretrained checkpoints and commercial APIs,
which leaves the attribution question open: is the gain due to the
MLLM backbone, or would \emph{any} encoder trained on the same
co-engagement pairs do as well? To answer this we train two
purpose-built non-LLM dual encoders, namely SigLIP2-large
(\texttt{patch16-384})~\cite{tschannen2025siglip2} and
CLIP ViT-L/14 (336px)~\cite{radford2021clip}, under a recipe that
is \emph{identical} to \method{} in every controllable respect.

\noindent\textbf{Matched protocol.}
All backbones use the same co-engagement pairs, the same symmetric
in-batch InfoNCE objective with the same number of cross-device
gathered negatives, the same $512$-dimensional output, the same
adaptation mechanism (frozen backbone $+$ LoRA $r{=}32$,
$\alpha{=}64$ on attention projections $+$ MLP projection head), the
same optimizer, learning rates, warm-up and cosine schedule, and the
same number of optimizer steps ($1$ full epoch). The dual encoders
reuse \method{}'s data loader, text renderer and loss implementation,
so the pairs, the field rendering and the negative pool are identical
by construction; the only differences are those intrinsic to the
backbone (its feature dimensionality, its image preprocessing, and
joint image--text attention versus dual-tower late fusion). Because a
$2$B MLLM does not fit the dual encoders' largest feasible batch on
$40$\,GB accelerators, we fix the negative count for \emph{all}
backbones at the largest value the MLLM supports, making compute and
negatives matched rather than favorable to \method{}.

\begin{table}[t]
  \caption{Matched backbones $\times$ serving modes. Within each
  serving mode, all backbones share the same co-engagement pairs,
  objective, negatives, output dimension, adaptation, schedule and
  number of steps, so the only difference down a column is the
  backbone. R@10 normalized to the \method{} (Qwen3-VL) multimodal
  entry. Best per serving mode in bold.}
  \label{tab:matched-backbones}
  \small
  \setlength{\tabcolsep}{2pt}
  \begin{tabular}{llccc}
    \toprule
    Backbone            & Type    & Multimodal         & Text-only         & Image-only        \\
    \midrule
    Qwen3-VL 2B (\method{}) & MLLM        & $\mathbf{100.0\%}$ & $\mathbf{89.0\%}$ & $68.8\%$          \\
    SigLIP2-large       & dual-enc.\  & $95.7\%$           & $60.1\%$          & $\mathbf{87.3\%}$ \\
    CLIP ViT-L/14       & dual-enc.\  & $94.5\%$           & $77.7\%$          & $73.8\%$          \\
    \bottomrule
  \end{tabular}
\end{table}

\noindent\textbf{Co-engagement supervision is the dominant
lever, but the backbone still matters.}
Table~\ref{tab:matched-backbones} shows that every backbone becomes a
usable co-engagement retriever once trained on the mined pairs: all
three land within $5.5$pp of each other on multimodal serving. The
spread across backbones is $5.8\%$ relative from best to worst,
against the $37.7\%$ relative gain that co-engagement fine-tuning
adds to the MLLM over its own pretrained checkpoint
(Table~\ref{tab:leaderboard}); we report pretrained baselines only for
the MLLM, so the rows here compare the three fine-tuned models.
The supervision, not the architecture,
produces the bulk of the gain: a useful and reassuring result for
practitioners, since it means the approach is portable. On top of that
shared gain the MLLM retains a small but \emph{statistically
significant} multimodal advantage: $4.3$pp over SigLIP2 and
$5.5$pp over CLIP (paired bootstrap and sign-flip permutation tests
over the $100$K test queries, $n{=}2000$ resamples, $p<0.001$ for both;
SigLIP2 versus CLIP is not significant). On text-only serving the
MLLM's margin is much larger ($89.0\%$ versus
$77.7\%$ and $60.1\%$),
consistent with product text being better served by a language model
than by a caption-oriented dual-encoder text tower.

\noindent\textbf{The deployed encoder is the same backbone under
different supervision.}
The production \texttt{I2I\_MM} control is itself a SigLIP2-large
encoder; what separates it from the SigLIP2 row of
Table~\ref{tab:matched-backbones} is that it is trained on
content-derived labels rather than on co-engagement pairs. The offline
split cannot price that difference fairly, because it is itself built
from co-engagement pairs and so favors any model trained on that
signal. The comparison that does
settle it is the online A/B of Section~\ref{sec:online}, where \method{}
replaces \texttt{I2I\_MM} on live traffic and lifts CTR by $+0.390\%$
and CVR by $+10.832\%$. Read together with the matched rows above,
which hold the supervision fixed and vary only the backbone, this
points to the change of supervision rather than a change of encoder
architecture as the source of the production gain.

\noindent\textbf{The MLLM keeps improving with training while the
dual encoders saturate.}
Evaluating intermediate checkpoints of the same three multimodal runs
(Table~\ref{tab:matched-curve}) shows the mechanism behind the
multimodal gap. \method{} improves monotonically across the epoch,
whereas both dual encoders plateau after roughly $40\%$ of it and then
flatten. The MLLM therefore has more headroom to align with the
engagement signal, and comparisons made at short training budgets
understate it: at $20\%$ of the epoch the gap to SigLIP2 is $2.3$pp,
and by the end it has widened to $4.3$pp.

\begin{table}[t]
  \caption{Multimodal R@10 versus optimizer steps for the matched
  runs of Table~\ref{tab:matched-backbones}, normalized to the final
  \method{} checkpoint. \method{} rises monotonically; the dual
  encoders plateau early.}
  \label{tab:matched-curve}
  \small
  \setlength{\tabcolsep}{4pt}
  \begin{tabular}{lccc}
    \toprule
    Steps (frac.\ of epoch) & \method{} & SigLIP2 & CLIP \\
    \midrule
    $20\%$                  & $97.1\%$  & $94.8\%$ & $93.6\%$ \\
    $40\%$                  & $98.4\%$  & $96.2\%$ & $94.1\%$ \\
    $60\%$                  & $99.5\%$  & $96.4\%$ & $93.7\%$ \\
    $80\%$                  & $99.6\%$  & $95.4\%$ & $94.4\%$ \\
    $100\%$                 & $\mathbf{100.0\%}$ & $95.7\%$ & $94.5\%$ \\
    \bottomrule
  \end{tabular}
\end{table}

\noindent\textbf{A dedicated vision encoder remains better for
image-only retrieval.}
The one mode where a dual encoder wins is pure image-only serving,
where SigLIP2 reaches $87.3\%$ against \method{}'s $68.8\%$. We read
this as a property of the readout rather than a claim about visual
understanding in general: SigLIP2's vision tower is contrastively
trained specifically to produce a standalone image embedding, whereas
the MLLM must express an image-only item through last-token pooling
over an image followed by empty text, which is outside its pretraining
regime. This is consistent with, and sharpens, the image-only gap
already reported in Section~\ref{sec:modality-flexibility}:
on image-only traffic the cost comes from the backbone rather than
from consolidation. Collapsing the two MLLM checkpoints into one costs
$1.8$pp (Table~\ref{tab:modality-flex}), whereas choosing an MLLM over
a contrastively-trained vision tower costs $18.5$pp, so a dedicated
visual encoder remains the better choice on that surface.

\subsection{Online experiments}
\label{sec:online}
We validate the offline findings in production. \method{} is
deployed in Snap DPA as a drop-in replacement for two content-based
encoders: the multimodal \texttt{I2I\_MM} encoder and the text-only \texttt{I2I\_TEXT}
encoder, which we do not evaluate on the offline split and therefore compare
against online only.
We run standard randomized A/B experiments with other serving
components unchanged (Table~\ref{tab:online}).

\begin{table}[t]
  \caption{Online A/B lifts for \method{} in Snap DPA. The
  first block reports the system-level effect on overall DPA traffic;
  the second reports per-surface lifts where \method{} replaces a
  specific deployed content encoder. Values are relative lifts. All
  three CVR lifts and the CTR lift against \texttt{I2I\_TEXT} are
  statistically significant at the $99\%$ confidence level.}
  \label{tab:online}
  \small
  \setlength{\tabcolsep}{4pt}
  \begin{tabular}{lcc}
    \toprule
    Comparison                          & $\Delta$ CTR      & $\Delta$ CVR       \\
    \midrule
    Overall DPA traffic           & $+0.211\%$  & $+1.911\%$   \\
    \midrule
    vs.\ \texttt{I2I\_MM} (MM)    & $+0.390\%$  & $+10.832\%$  \\
    vs.\ \texttt{I2I\_TEXT} (text) & $+18.958\%$ & $+13.12\%$  \\
    \bottomrule
  \end{tabular}
\end{table}

\noindent\textbf{CVR improves on every comparison, while CTR gains
concentrate on the text surface.}
At the system level, \method{} raises overall DPA CVR by $+1.911\%$;
the overall CTR change ($+0.211\%$) is positive but does not reach
significance. These aggregate figures are measured on
total DPA traffic, of which the two replaced encoders serve a subset,
so the per-surface lifts below do not translate proportionally to the
system level. Restricted to the surfaces it takes over, the effect is
sharper.
Against \texttt{I2I\_MM}, the measurable gain is entirely in
conversion: CVR rises $+10.832\%$ while the CTR change ($+0.390\%$) is
not significant, indicating that co-engagement supervision surfaces
more conversion-relevant candidates without changing which items look
clickable.
Against \texttt{I2I\_TEXT}, both metrics move sharply
($+18.958\%$ CTR and $+13.12\%$ CVR), and here the ordering reverses
(the CTR gain exceeds the CVR gain), consistent with the offline
finding that adding image understanding to a
text-only retriever recovers candidates the control did not retrieve at
all.

\section{Conclusion}
\label{sec:conclusion}

We presented \method{}, a co-engagement-aware framework for
training modality-flexible item embeddings for industrial I2I
retrieval. \method{} uses LLM/MLLM backbones to represent item
images and metadata in a shared embedding space, then aligns that
space with production DPA co-engagement signals using symmetric
in-batch InfoNCE. The resulting embeddings fit the existing ANN
serving path and support multimodal, text-only, and image-only
retrieval modes.

On Snap DPA, the Qwen3-VL-Embedding 2B instantiation outperforms
pretrained open-source backbones and the strongest commercial
multimodal embedding model on offline Recall@10, serves text-only
retrieval at
near-parity with a dedicated text encoder from the same checkpoint,
and benefits more from training-data scale than from filtering
co-engagement pairs down to high-intent events. A matched-recipe
comparison against purpose-built dual encoders locates the gain: the
co-engagement supervision accounts for most of it and transfers across
backbone families, while the MLLM adds a modest multimodal and a large
text-only advantage, with image-only retrieval the exception. In
production A/B tests \method{} delivers $+0.390\%$ CTR /
$+10.832\%$ CVR over the multimodal control, $+18.958\%$ CTR /
$+13.12\%$ CVR over the text control, and $+0.211\%$ CTR /
$+1.911\%$ CVR on overall DPA traffic; it has since launched in place
of both encoders. The lesson for practitioners is that the behavioral
supervision, not the backbone family, buys most of the quality, and
that the value of an MLLM is one checkpoint serving the multimodal and
text surfaces competitively.

%% Acknowledgements: optional. KDD ADS review is single-blind, so these
%% may be included in the submission itself.

\bibliographystyle{ACM-Reference-Format}
\bibliography{main}

\end{document}